\documentclass[12pt]{article}

\usepackage{latexsym}

\usepackage{graphicx}

\begin{document}


\title{Magnetized static black Saturn
}

\author{
     Stoytcho S. Yazadjiev \thanks{E-mail: yazadj@theorie.physik.uni-goettingen.de; yazad@phys.uni-sofia.bg}\\
{\footnotesize  Institut f\"ur Theoretische Physik, Universit\"at G\"ottingen,}\\
{\footnotesize  Friedrich-Hund-Platz 1, D-37077 G\"ottingen, Germany}\\
{\footnotesize and} \\
{\footnotesize  Department of Theoretical Physics,
                Faculty of Physics, Sofia University,}\\
{\footnotesize  5 James Bourchier Boulevard, Sofia~1164, Bulgaria }\\
}

\date{}

\maketitle

\begin{abstract}
We construct a new static  solution to the 5D Einstein-Maxwell equations describing a static black hole
surrounded by a non-rotating dipole black ring. The configuration is kept in equilibrium by an external magnetic field interacting with
the dipole charge of the black ring. The  properties of the black Saturn-like configuration are studied and the basic physical quantities
are calculated. The solution demonstrates 2-fold continuous non-uniqueness of the 5D magnetized static neutral black objects for fixed total
mass and Melvin background.
\end{abstract}


\sloppy

\section{Introduction}
In the last years we observe rapidly growing interest in the higher dimensional black holes.
Only in five dimensions, however, we have exact solutions
describing more non-trivial black objects as black rings, black Saturns, concentric and
orthogonal black di-rings \cite{ER2}-\cite{ElvangRodriguez}.
Some basic steps towards the classification of the five dimensional asymptotically flat
black solutions were also made \cite{HolandsYazadjiev1},\cite{HolandsYazadjiev2}.
Nevertheless, our knowledge of the 5D black objects and their physical properties is rather incomplete.
Let us mention some of the open problems. Are there black holes and black rings with rod structure different from
the standard one? Are there black Saturns and black multi-rings for which the balancing mechanism is different, or partially different
from the centrifugal repulsion? In particular, are there static charged black Saturns and multi-rings where the electric charge
balances the gravitational attraction?
In this situation  the exact solutions
remain almost the only way to obtain new valuable and reliable knowledge of the black objects and their physical properties.
Most of the known exact solutions describing black objects with nontrivial horizon topology are asymptotically flat.
There are, however, asymptotically non-flat
black solutions which are also interesting from physical point of view. As an example we can mention the so-called magnetized
solutions describing various kinds of objects immersed in a self-gravitating magnetic field \cite{Ortaggio},\cite{Yazadjiev2}.

The aim of this paper is to present a new static solution describing a non-rotating black hole surrounded by a non-rotating dipole
black ring in the 5D Einstein-Maxwell (EM) gravity. The described solution will be called magnetized, static, dipole black Saturn.
In contrast to the case
of the known black Saturn and di-ring solutions \cite{ElvangFigueras}--\cite{ElvangRodriguez}
where the black configurations are balanced by  the rotation of the black ring(s), in our case the balancing mechanism
is quite  different.  The configuration is kept in equilibrium by a self-gravitating external magnetic field interacting
with the dipole charge. This interaction, namely, provides the additional force which can balance the  configuration.

\section{Static dipole black Saturn in external magnetic field }
The 5D EM gravity is described by the field equations
\begin{eqnarray}\label{EMFE}
&&R_{\mu\nu} = {1\over 2} \left(F_{\mu\lambda}F_{\nu}^{\,\lambda}
 - {1\over 6} F_{\sigma\lambda}F^{\sigma\lambda} g_{\mu\nu}\right),  \\
&&\nabla_{\mu} F^{\mu\nu} = 0  =\nabla_{[\mu}F_{\nu\sigma]}\nonumber.
\end{eqnarray}

One of the ways to construct the magnetized static dipole black Saturn solution  to the field equations (\ref{EMFE}) is
to  magnetize first the stationary dipole black Saturn of \cite{Yazadjiev} and then take the static limit. The disadvantage of this
approach is the fact that the explicit form of the gauge potential is not known for the stationary dipole black Saturn. That is why,
we prefer to use
more constructive approach in order to present the solution in completely explicit form. Namely, using the solution generating method
of \cite{HolandsYazadjiev1} we generate first the configuration\footnote{Although it is clear it is worth noting that this configuration
suffers from the presence of conical singularities.} of a static black hole surrounded by a static dipole black ring.
Then we magnetize this static solution by applying the Harrison transformation \cite{Ortaggio},\cite{Yazadjiev2}.
As a result of the procedure described we obtain the following explicit solution
\begin{eqnarray}
&&ds^2 = {\Lambda \over S}\left[ {\cal G}_{t} dt^2 +  Y^3 e^{2K} (d\rho^2 + dz^2) + {\cal G}_{\psi}d\psi^2\right]
+ {S^2\over \Lambda^{2}} {\cal G}_{\phi}d\phi^2 ,\\
&&A_{\phi}= \Lambda^{-1}\left[{\cal U}_{\phi} + {\sqrt{3}b\over 2}\left(S^2 {\cal G}_{\phi} + {4\over 3}{\cal U}_{\phi}^2\right)    \right]
\end{eqnarray}
where
\begin{eqnarray}
\Lambda= {1\over 3}B^2S^2{\cal G}_{\phi} + \left(1 + {2\over 3}B {\cal U}_{\phi} \right)^2
\end{eqnarray}
and the functions ${\cal G}_{t}$, ${\cal G}_{\psi}$, ${\cal G}_{\phi}$, $e^{2K}$, $S$, $Y$,  and ${\cal U}_{\phi}$ are given by
\begin{eqnarray}
&&{\cal G}_{t} = - {\mu_{1}\mu_{3}\over \mu_{2}\mu_{4} },\,\,\,\,\,\, {\cal G}_{\psi} = \mu_{2}, \,\,\,\,\,\,
{\cal G}_{\phi}= {\rho^2\mu_{4}\over \mu_{1}\mu_{3}} ,\\
&&e^{2K}= e^{2K_{0}}{\mu_{2}(\rho^2 + \mu_{1}\mu_{2})(\rho^2 + \mu_{1}\mu_{4})^2 (\rho^2 + \mu_{2}\mu_{3})(\rho^2 + \mu_{3}\mu_{4})^2 \over (\rho^2 + \mu_{1}\mu_{3})^2 (\rho^2 + \mu_{2}\mu_{4}) \prod_{i=1}^{4} (\rho^2 + \mu^2_{i}) },\\
&&S={(1-\omega)(1-\nu)R_{1} + (1+\omega)(1 + \nu)R_{4} + 2(\omega + \nu)R_{3} + \omega(1-\nu^2){\cal R}^2 \over
(1-\omega)(1-\nu)R_{1} + (1+\omega)(1 + \nu)R_{4} + 2(\omega + \nu)R_{3} - \omega(1-\nu^2){\cal R}^2 },\\
&& Y = {(1-\omega)(1-\nu)R_{1} + (1+\omega)(1 + \nu)R_{4} + 2(\omega + \nu)R_{3} + \omega(1-\nu^2){\cal R}^2 \over
[ (1-\nu)R_{1} + (1+\nu)R_{4} + 2\nu R_{3} ] },\\
&&{\cal U}_{\phi}= {2\sqrt{3}(1-\nu){\cal R}C(\nu, -\omega)\left[R_{1} - R_{3} + {1\over 2}(1+\nu){\cal R}^2\right]\over
(1-\omega)(1-\nu)R_{1} + (1+ \omega)(1+\nu)R_{4} + 2(\nu +\omega)R_{3} - \omega (1-\nu^2){\cal R}^2 }.
\end{eqnarray}

The functions $R_{i}$ and $\mu_{i}$ are defined by
\begin{eqnarray}
R_{i}=\sqrt{\rho^2 + (z-a_{i})^2}, \,\,\,\,\,\,\,\,\, \mu_{i}= R_{i} - (z-a_{i}).
\end{eqnarray}

In all the above expressions $a_{i}$, $\omega$, $\nu$,  ${\cal R}$ and $B$ are real constants with
\begin{eqnarray}
a_{1}= -{\nu\over 2}{\cal R}^2 ,  \,\,\, a_{2}={\zeta\over 2} {\cal R}^2,\,\,\,   a_{3}= {1\over 2}{\cal R}^2, \,\,\,\, a_{4}= {\nu\over 2}{\cal R}^2,
\,\,\, C(\nu, -\omega)=\sqrt{\omega(\omega + \nu){1-\omega\over 1+\omega }}
\end{eqnarray}
where\footnote{In other words we have $a_{1}\le a_{4} < a_{3}\le a_{2}$.}
\begin{eqnarray}
0\le \omega<1, \,\,\,  0\le \nu <1 ,\,\,\,\, 1\le \zeta  .
\end{eqnarray}

The solution depends on six independent parameters -- $\omega, \nu, \zeta, {\cal R}$, $B$ $(B>0)$ and $e^{2K_{0}}$.
The parameter $B$, as we will see,  is  the asymptotic strength of the external magnetic field.
\section{Analysis of the  solution}

\subsection{Asymptotic behavior}

In order to study the asymptotic behaviour of the solution  we  introduce
the asymptotic coordinates $r$ and $\theta$ defined by
\begin{eqnarray}
\rho = {1\over 2}r^2 \sin 2\theta ,\,\,\,  z = {1\over 2}r^2 \cos 2\theta.
\end{eqnarray}
Then in the asymptotic limit $r\to \infty$  we find
\begin{eqnarray}
&&{\cal G}_{t} \approx -1 + \left(2\nu + \zeta -1 \right){{\cal R}^2\over r^2},\\
&&{\cal G}_{\psi} \approx r^2 \sin^2\theta,\\
&&{\cal G}_{\phi} \approx r^2\cos^2\theta,\\
&& S \approx 1 + {2\omega (1-\nu)\over 1+ \omega }{{\cal R}^2\over r^2},\\
&& Y \approx (1+\omega)\left[1 + {2\omega (1-\nu)\over 1+ \omega }{{\cal R}^2\over r^2} \sin^2\theta\right],\\
&&e^{2K}\approx {e^{2K_{0}} \over r^2 },\\
&&{\cal U}_{\phi}\approx {2\sqrt{3}(1-\nu){\cal R}C(\nu,-\omega)\over 1+ \omega } {{\cal R}^2\over r^2}\cos^2\theta.
\end{eqnarray}
Therefore the asymptotic form of the solution is the following
\begin{eqnarray}
ds^2 \approx &&\left(1 + {1\over 3}B^2 r^2\cos^2\theta\right) \left[-dt^2 + (1+ \omega)^3 e^{2K_{0}} (dr^2 + r^2d\theta^2) + r^2\sin^2\theta d\psi^2 \right]
\nonumber \\&&+ {r^2\cos^2\theta d\phi^2 \over (1 + {1\over 3}B^2 r^2\cos^2\theta)^2} ,\\
A_{\phi}\approx &&{1\over 2}B {r^2\cos^2\theta \over (1 + {1\over 3}B^2 r^2\cos^2\theta)}.
\end{eqnarray}

With the choice
\begin{eqnarray}\label{MelvinC}
e^{2K_{0}}=  (1+\omega)^{-3}
\end{eqnarray}
the solution  approaches the 5D Melvin universe with external axial magnetic field $B$.
From now on we adopt that the constant $e^{2K_{0}}$ is given by  (\ref{MelvinC}).

\subsection{Rod structure}
The rod structure is the following.

(i) Semi-infinite rod $(-\infty,a_{1}]$ and the finite rod $[a_{4},a_{3}]$ corresponding to the $\phi\phi$-part of the metric

In order to cure the conical singularities at the location of the semi-infinite rod the coordinate $\phi$ must have a period
\begin{equation}
\Delta \phi = 2\pi \lim_{\rho\to 0}\sqrt{\rho^2 g_{\rho\rho}\over g_{\phi\phi}}
\end{equation}
which gives
\begin{equation}\label{PHIRC1}
\Delta \phi = 2\pi (1+ \omega)^{3/2} e^{K_{0}}\Lambda^{3/2}(\rho=0)= 2\pi .
\end{equation}
Respectively, for the finite rod $[a_{4},a_{3}]$ we find
\begin{equation}
\Delta \phi = 2\pi \lim_{\rho\to 0}\sqrt{\rho^2 g_{\rho\rho}\over g_{\phi\phi}}= 2\pi (1-\omega)^{3/2} e^{K_{0}} \left(1-\nu\over 1+ \nu \right)
\sqrt{\zeta + \nu \over \zeta -\nu } \Lambda^{3/2}(\rho=0).
\end{equation}
For the rod  $[a_{4},a_{3}]$ we have
\begin{eqnarray}
\Lambda(\rho=0)=\left(1 + {2\over 3}B{\cal Q} \right)^{2}
\end{eqnarray}
and therefore we can write
\begin{equation} \label{PHIRC2}
\Delta \phi = 2\pi \left(1-\omega\over 1 +\omega\right)^{3/2} \left(1-\nu\over 1+ \nu \right)
\sqrt{\zeta + \nu \over \zeta -\nu } \left(1 + {2\over 3}B{\cal Q} \right)^{3}.
\end{equation}
Here ${\cal Q}$ is the dipole charge of the non-magnetized dipole black ring. The explicit formula for ${\cal Q}$ is presented
below (see eq. (\ref{NONMAGDC})).

(ii) Semi-infinite rod $[a_{2}, \infty)$  corresponding to the $\psi\psi$-part of the metric

The regularity condition here gives
\begin{eqnarray}
\Delta \psi = 2\pi \lim_{\rho\to 0}\sqrt{\rho^2 g_{\rho\rho}\over g_{\psi\psi}}= 2\pi (1+ \omega)^{3/2} e^{K_{0}}=2\pi.
\end{eqnarray}

In order to find the balancing condition we must impose the r.h.s of (\ref{PHIRC2}) to be equal to that of (\ref{PHIRC1}). In other words
the balancing condition is
\begin{eqnarray}\label{BALANCEC}
\left(1-\omega\over 1 +\omega\right)^{3/2} \left(1-\nu\over 1+ \nu \right)
\sqrt{\zeta + \nu \over \zeta -\nu } \left(1 + {2\over 3}B{\cal Q} \right)^{3}=1.
\end{eqnarray}

The eq.(\ref{BALANCEC}) can be solved to determine  $B$ as a function  of the remaining parameters.
In other words the external magnetic field can always be chosen such as to cancel the conical singularity and
to support the static Saturn in equilibrium. This is possible because of the coupling between the external magnetic filed $B$ and
the dipole charge ${\cal Q}$. This coupling provides the additional force which compensates the gravitational attraction
between the black hole and the black ring and the magnetic self-interaction of the dipole black ring.

\subsection{Horizons}

\subsubsection{Black ring horizon}

The black ring horizon is located at $\rho=0$ for $a_{1}\le z\le a_{4}$. The metric of the spacial cross section
of the black ring horizon is given by
\begin{eqnarray}
ds^2_{BR}= {\Lambda_{BR}\over S_{BR}}\left[Y^3_{BR}e^{2K_{BR}}dz^2 + {\cal G}^{BR}_{\psi}d\psi^2 \right]
+ {S^2_{BR}\over \Lambda^2_{BR}}{\cal G}^{BR}_{\phi}d\phi^2
\end{eqnarray}
where
\begin{eqnarray}
&&S_{BR}= {{1\over 2}{\cal R}^2 - z\over {1\over 2}{\cal R}^2{\nu (1+ \nu\omega)\over \omega + \nu} - z},\\
&& Y_{BR}= {\omega + \nu \over \nu},\\
&&e^{2K_{BR}}= {\cal R}^2{(\zeta + \nu)\nu^2\over (1+\nu)^2 (1+ \omega)^3 } {({1\over 2}{\cal R}^2 - z)
\over (z + {\nu\over 2}{\cal R}^2) ({\nu\over 2}{\cal R}^2-z)({\zeta\over 2}{\cal R}^2 -z)},\\
&&{\cal G}^{BR}_{\psi}=2\left({\zeta\over 2}{\cal R}^2-z\right), \\
&&{\cal G}^{BR}_{\phi}= 2 {(z + {\nu\over 2}{\cal R}^2)({\nu\over 2}{\cal R}^2 - z) \over ({1\over 2}{\cal R}^2 - z)},\\
&&\Lambda_{BR}= b^2 S^{2}_{BR} {\cal G}^{BR}_{\phi} + \left(1 + {2b\over 3}{\cal U}^{BR}_{\phi}\right)^2 ,\\
&&{\cal U}^{BR}_{\phi} = {\sqrt{3}(1-\nu){\cal R}C(\nu,-\omega)\over (\omega + \nu)} {(z+{\nu\over 2}{\cal R}^2) \over(
{\nu(1 + \omega \nu)\over 2(\omega + \nu)}{\cal R}^2 -z)}.
\end{eqnarray}

Since the orbits of $\phi$ shrink to zero size at $z=a_{1}=-{\nu\over2 }{\cal R}^2$ and $z=a_{4}={\nu\over 2}{\cal R}^2$
and the orbits of $\psi$ do not shrink to zero size anywhere (for $a_{1}\le z\le a_{4}$ ), the topology of the horizon is $S^2\times S^1$.
Metrically, however, the horizon is distorted by the external magnetic field and the gravitational
attraction by the black hole. The area of the horizon can be found by direct calculations
\begin{eqnarray}
{\cal A}_{BR}= 8\pi^2 {{\cal R}^3\over (1+ \nu) } \sqrt{\nu(\zeta + \nu) \left({\omega + \nu \over 1 + \omega}\right)^3}.
\end{eqnarray}

The temperature of the black ring horizon can be obtained form the surface gravity and it is given by
\begin{eqnarray}
T_{BR}= {1+\nu\over 4\pi {\cal R} } \sqrt{{\nu \over \zeta + \nu}\left({1+ \omega \over \omega + \nu }\right)^3}.
\end{eqnarray}
The same result is obtained by using the euclidian approach.

Let us note that the area and the temperature of the black ring horizon are not affected by the external magnetic field
and they coincide with the corresponding quantities of  the non-magnetized solution. This is in complete agreement with
what was observed for other  the magnetized static black solutions -- the external
magnetic field does not affect the thermodynamics \cite{Radu},\cite{Ortaggio},\cite{Yazadjiev2}.

\subsubsection{Black hole horizon}

The black ring horizon is located at $\rho=0$ for $a_{3}\le z\le a_{2}$. The metric of the spacial cross section
of the black hole horizon is
\begin{eqnarray}
ds^2_{BH}= {\Lambda_{BH}\over S_{BH}}\left[Y^3_{BH}e^{2K_{BH}}dz^2 + {\cal G}^{BH}_{\psi}d\psi^2 \right]
+ {S^2_{BH}\over \Lambda^2_{BH}}{\cal G}^{BH}_{\phi}d\phi^2
\end{eqnarray}
and the metric functions are given as follows
\begin{eqnarray}
&&S_{BH}=  {z- {\nu\over 2}{\cal R}^2 \over z - {2\omega+ \nu(1-\omega)\over 2(1+ \omega)}{\cal R}^2},\\
&&Y_{BH}= 1 + \omega ,\\
&&e^{2K_{BH}}= {\cal R}^2 {(\zeta + \nu)(\zeta-1)\over 4(1+ \omega)^3(\zeta-\nu)} {(z- {\nu\over 2}{\cal R}^2)\over
(z+ {\nu\over 2}{\cal R}^2)({\zeta\over 2}{\cal R}^2 -z)(z- {1\over 2}{\cal R}^2)},\\
&&{\cal G}^{BH}_{\psi}= 2\left({\zeta\over 2}{\cal R}^2 -z\right),\\
&&{\cal G}^{BH}_{\phi}=2 {(z + {\nu\over 2}{\cal R}^2)(z- {1\over 2}{\cal R}^2) \over (z - {\nu\over 2}{\cal R}^2) },\\
&&\Lambda_{BH}= b^2 S^{2}_{BH} {\cal G}^{BH}_{\phi} + \left(1 + {2b\over 3}{\cal U}^{BH}_{\phi}\right)^2 ,\\
&&{\cal U}^{BH}_{\phi} = {\sqrt{3}(1-\nu){\cal R}^3 C(\nu,-\omega)\over (1+ \omega)(z - {2\omega + \nu(1-\omega)\over 2(1+\omega) }{\cal R}^2) }.
\end{eqnarray}

The orbits of $\phi$ shrink to zero size at $z=a_{3}={1\over 2}{\cal R}^2$ and the orbits of $\psi$ at $z=a_{2}={\zeta\over 2 }{\cal R}^2$. Therefore
the topology of the horizon is $S^3$. The 3-sphere of the horizon is metrically distorted by the self-gravitating external magnetic field and
the gravitational attraction of the black ring.

The  area of the black hole horizon is
\begin{eqnarray}
{\cal A}_{BH} = 2\pi^2 {\cal R}^3 \sqrt{{(\zeta +\nu)(\zeta-1)^3\over (\zeta-\nu)}}
\end{eqnarray}
and temperature of the black hole is given by
\begin{eqnarray}
T_{BH}= {1\over 2\pi {\cal R}} \sqrt{{\zeta-\nu\over (\zeta + \nu)(\zeta -1)}}.
\end{eqnarray}

The area and the temperature of the black hole are not affected by the external magnetic field.

\subsection{Mass and dipole charge }

The dipole charge is defined by
\begin{eqnarray}
{\cal Q}_{M}= {1\over 2\pi} \oint_{S^2} F
\end{eqnarray}
where $S^2$ is the two-dimensional sphere of the black ring horizon. The subscript "M" means that the dipole charge is for the magnetized solution.
By straightforward calculations
we find that the dipole charge is given by
\begin{eqnarray}
{\cal Q}_{M}= {{\cal Q}\over 1 + {2\over 3}{B}{\cal Q}}
\end{eqnarray}
where
\begin{eqnarray}\label{NONMAGDC}
{\cal Q}= 2\sqrt{3}{\cal R}{C(\nu,-\omega) \over 1-\omega}
\end{eqnarray}
is the dipole charge of the non-magnetized dipole black ring.

In order to compute the mass of our black configuration we use the quasi-local formalism.
Following this approach we have to choose appropriate reference background. The most natural reference background is the Melvin
space-time. With this choice following the detailed calculations presented in \cite{Yazadjiev2} we find that the mass of the static black Saturn
is given by
\begin{eqnarray}
M = {3\pi\over 8}{\cal R}^2 \left[\zeta + {2\nu -1 + \omega \over 1 + \omega } \right].
\end{eqnarray}

\subsection{Limits of the solution }
The solution has two natural non-singular limits. By removing the black hole from the configuration (i.e. by setting $\zeta=1$)
we obtain the magnetized static black ring \cite{Ortaggio}. Removing the dipole black ring from the configuration we obtain the magnetized
black hole. This is achieved by setting first $\omega=0$ and then taking the limit $\nu\to 0$.

\section{Discussion}
In this work we have constructed a new solution of the 5D EM equations describing a static black hole surrounded by a static dipole black ring.
This Saturn-like configuration is balanced by the force yielded by the interaction of the dipole charge with an external self-gravitating magnetic
field.
The balanced solution depends on four independent parameters. Only one of them, the mass $M$, is a conserved quantity. Therefore our solution
exhibits 3-fold continuous non-uniqueness. If we consider the solution in fixed Melvin background (i.e. keeping $B$ fixed) which is the most
interesting case from physical point of view, the solution then exhibits 2-fold continuous non-uniqueness. Therefore our solution
demonstrates 2-fold continuous non-uniqueness of the 5D static, magnetized (neutral) black objects in fixed Melvin background.
Let us denote by ${\cal A}^{Saturn}$ the total area of the magnetized static black Saturn and by ${\cal A}^{hole}$ and ${\cal A}^{ring}$
the area of the single magnetized static black hole and black ring. It can be shown  that
\begin{eqnarray}
0< {\cal A}^{ring} < {\cal A}^{Saturn}<{\cal A}^{hole}
\end{eqnarray}
for black configurations with the same total mass and immersed in the same  Melvin background. Moreover, the magnetized static black
Saturns swept the entire interval $(0, {\cal A}^{hole})$.

\section*{Acknowledgements}
The author would like to thank the Alexander von Humboldt Foundation for a stipend, and
the Institut f\" ur Theoretische Physik G\" ottingen for its kind hospitality. The partial support by the
Bulgarian National Science Fund under Grant MUF04/05 (MU 408) and VUF-201/06  is also acknowledged.

\end{document}